\documentclass[prb,twocolumn,floatfix]{revtex4}

\usepackage{amsmath}
\usepackage{epsfig}

\renewcommand{\vec}[1]{{\rm\bf #1}}
\newcommand{\Tr}{\mathop{\mathrm{Tr}}}

\newcommand{\cc}{\mathop{\rm c.\,c.}}

\newcommand{\sign}{\mathop{\mathrm{sign}}}

\newcommand{\ep}{\epsilon}

\newcommand{\unitmatrix}{\openone}
\newcommand{\rep}{\mathrm{r}}

\newcommand{\range}{R}
\newcommand{\cut}{r_0}
\newcommand{\valley}{\kappa}
\newcommand{\chirality}{s}
\newcommand{\diag}{W}
\newcommand{\timerev}{U_t}
\newcommand{\Lmat}{L}

\begin{document}

\title{Resonant low-energy electron scattering on
short-range impurities in graphene}
\author{D.~M.~Basko}\email{basko@sissa.it}
\affiliation{International School of Advanced Studies (SISSA), via
Beirut~2-4, 34014 Trieste, Italy}

\begin{abstract}
Resonant scattering of electrons with low energies (as compared to
the bandwidth) on a single neutral short-range impurity in graphene
is analyzed theoretically, taking into account the valley degeneracy.
Resonances dramatically increase the scattering cross-section and
introduce a strong energy dependence. Analysis of the tight-binding
model shows that resonant scattering is typical for generic impurities
as long as they are sufficiently strong
(the potential is of the order of the electron bandwidth or higher).
\end{abstract}

\maketitle

\section{Introduction}

Electron transport in graphene is a subject of intense study,
both theoretical and experimental, since the very discovery of
this material in 2004.\cite{Novoselov} In general, electron
transport is determined by competition of different scattering
mechanisms, both inelastic (e.~g., electron-phonon) and
elastic (static defects). Elastic scattering is dominant at
sufficiently low temperatures.

Different kinds of crystal imperfections can cause elastic
electron scattering in graphene: mesoscopic corrugations of the
graphene sheet (ripples)\cite{ripples} producing perturbations
smooth on the atomic scale, charged impurities producing long-range
Coulomb fields, dislocations producing long-range strain fields,
or short-range neutral impurities. While the first
three types seem to be more important for the transport in the
clean graphene, it is probably the fourth one that can be controlled.
In Refs.~\onlinecite{Gilje2007,Gomez2007} graphene oxide was
chemically reduced to normal graphene; in Ref.~\onlinecite{Echtermeyer}
hydrogen and hydroxil groups were deposited on the graphene sheet
in a controlled and reversible manner; in all cases resistance
changed by several orders of magnitude.  Assuming that attachment
of a chemical group to a carbon atom in graphene changes the
hybridization of its electronic orbitals from $sp^2$ to $sp^3$, one
can view such group as a neutral short-range impurity.

Short-range impurities in carbon nanotubes have been studied even
before the single-layer graphene was obtained in the laboratory.%
\cite{Ando1998,Cohen2000,Song2002,Baranger2003,McCann2005}
Short-range impurities have been shown to modify local electronic
properties of graphene, such as the local density of states\cite{%
Pereira2006,Pogorelov,Peres2006,Loktev,Lichtenstein,Peres2007,Bena}
or local magnetic moment,\cite{Peres2006,Yazyev} and to induce
Friedel oscillations in doped samples.\cite{CheianovFalko}
The present work is dedicated to the problem of electron scattering
on a single short-range impurity whose size~$\range$ is assumed
to be of the order of the interatomic distance~$a$ (the C--C bond length),
and the electron energy~$\ep$ is assumed to be much smaller than the
energy scale set by the potential, $v/\range$ ($v$~is the electron
velocity at the Dirac point, so that $v/\range$ is of the order of the
electronic bandwidth). The main focus is the case of a strong impurity
so that the results for electron scattering obtained in the first Born
approximation\cite{ShonAndo,SuzuuraAndo,Khveshchenko,%
McCann,NomuraMacDonald,KechedzhiFalko} are not expected to
be applicable. Instead, we are going to exploit the smallness
$|\ep|\range/v\ll{1}$.
For particles with parabolic spectrum such low-energy scattering is
characterized by a single constant of the dimensionality of length
(the scattering length), determined from the solution of the
Schr\"odinger equation at zero energy. In graphene, due to the
degeneracy of the spectrum at the Dirac point, more than one length
is needed to characterize a scatterer.\cite{AleinerEfetov}

The common intuition is that for a strong enough scatterer the
typical value of scattering lengths $l\sim\range$, yielding the
cross-section $\sigma\sim{p}l^2\ll{a}$ (here $p=|\ep|/v$ is the
electron momentum counted from the Dirac point). The exception
to this is the case of resonant scattering when the potential has a
(quasi-)bound state with small energy, then one of the scattering
lengths becomes of the order of the size of this state. The main
motivation for the present study is that for the Dirac spectrum the
exception becomes a rule: a vacancy (which can be viewed as
the limit of an infinitely strong scatterer) introduces a bound state
exactly at the Dirac point,\cite{Pereira2006}
so one of the scattering lengths diverges. As a consequence, the
scattering cross-section diverges as the electron energy approaches
the Dirac point. This divergence corresponds to that found in the
scattering off vacancies\cite{Peres2006,Stauber2007}
and in the unitary limit of potential impurities.\cite{Mirlin}

It is hard to introduce a
real vacancy in graphene, however, if the $\pi$-orbital of some
carbon atom is very tightly bound, this atom acts effectively as a
vacancy for the rest of the $\pi$-electrons in the crystal. In
particular, impurities introduced
electrochemically\cite{Echtermeyer,Gilje2007,Gomez2007} are
likely to act as strong scatterers, and thus are unlikely to be
described by the first Born approximation.
This expectation is supported by
density-functional theory (DFT) calculations for
graphane\cite{graphane} (a hypothetical material obtained from
graphene by attaching a hydrogen atom to each carbon): the binding
energy was obtained to be about 7~eV per hydrogen atom. For
such scatterers the scattering length should be large, thus motivating
the present study.

The main framework of this study is the general scattering
theory\cite{Newton} modified for the 2D Dirac equation%
\cite{Mele1984,Katsnelson2007,Novikov2007}
taking into account the valley degeneracy.
In Sec.~\ref{sec:scattering} we show that for a general short-range
scatterer all the information necessary to determine the cross-section
up to corrections of the order $(p\range)^2$ is encoded in a $4\times{4}$
matrix~$\Lmat$ and a constant~$\cut$ (all having the
dimensionality of length). The terms of the order $(p\range)^2$ and
higher cannot be studied using the Dirac equation, as the Dirac
hamiltonian itself is the leading term in the expansion of a microscopic
hamiltonian in the parameter~$pa$ (and $\range\sim{a}$ is always
assumed here). The matrix~$\Lmat$ (i)~can be obtained from the solution
of the microscopic Schr\"odinger equation (e.~g., an {\it ab initio} calculation)
for electrons in the graphene crystal at zero energy and with an appropriate
asymptotics; (ii)~is hermitian and invariant with respect to the time reversal,
so it depends on 10 real parameters; (iii)~its four eigenvalues $l_1,\ldots,l_4$
play the role of the scattering lengths. Divergence of one or several of these eigenvalues
signals the existence of a localized solution at zero energy.
For the parameter~$\cut$ we have (i)~$\cut\sim\range$; (ii)~the dependence
of the scattering amplitude on~$\cut$ is weak (logarithmic); (iii)~the
exact value of~$\cut$ cannot be extracted from the zero-energy solutions
only, wave functions at low but finite energies have to be considered in order
to determine it. When $L$~and~$\cut$ are known, the low-energy scattering
$T$-matrix is given by Eq.~(\ref{Tmatrixres=}), which covers both Born
limit [$T(\ep)$~is $\ep$-independent, the cross-section $\sigma\propto\ep$]
and the unitary limit [$T(\ep)\propto1/(\ep\ln\ep)$,
$\sigma\propto{1}/|\ep\ln^2\ep|$], as well as the crossover between them
for an impurity of a large but finite strength.

\begin{figure}
\includegraphics[width=8cm]{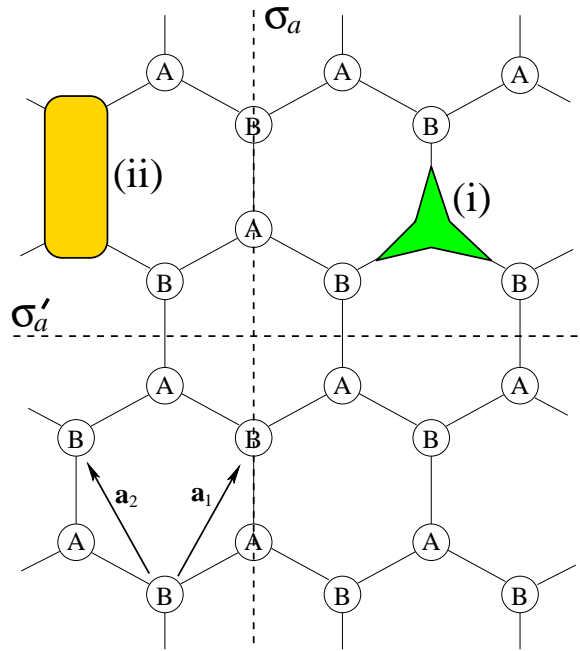}
\caption{\label{fig:lattice} (Color online). The honeycomb lattice with
two atoms ($A$~and~$B$) per unit cell and the elementary translation
vectors $\vec{a}_1$ and $\vec{a}_2$. The two inequivalent reflection
planes $\sigma_a$ and $\sigma_a'$ are shown. (i)~and~(ii) are schematic
representations of site-like and bond-like impurities with symmetries
$C_{3v}$ and $C_{2v}$, respectively.}
\end{figure}

In Sec.~\ref{sec:special} we consider two examples of impurities with
special symmetries, and see how these symmetries manifest themselves
in the scattering (i.~e., how they restrict the form of the
matrix~$\Lmat$). The first example (the site-like impurity) is an impurity
localized around one of the carbon atoms and preserving its $C_{3v}$
symmetry ($C_{3v}$~consists of three-fold rotations and reflections in three
planes perpendicular to the crystal plane); it is natural to assume that this
would be the case for a hydrogen atom bound to a carbon. The second kind
(the bond-like impurity) involves two neighboring carbon atoms and the bond
between them and has the symmetry~$C_{2v}$. This could be the case
for an oxigen atom bound to two carbon atoms. The two kinds of impurities,
described above, are schematically shown in Fig.~\ref{fig:lattice}. Of
course, a generic impurity is not going to have any symmetry at all.

Since graphene crystal is symmetric, impurities of the same kind
can occur in different locations and with different orientations with
equal probability, if these can be related to each other by a crystal
symmetry operation. For example, the site-like impurity, located
on an $A$~atom in Fig.~\ref{fig:lattice}, can reside on a $B$~atom
with the same probability; the bond-like impurity can have one
of the three different orientations, rotated by $2\pi/3$ with respect
to each other. Although equivalent from the crystal symmetry
point of view, such impurities will have different
$\Lmat$~matrices. If one is not looking at effects of coherent
scattering off several impurities, the cross-section can be averaged
over such equally probable impurity configurations. This procedure
is described in Sec.~\ref{sec:averaging}.

In Sec.~\ref{sec:tight} we perform explicit calculations in the
tight-binding model as an example of microscopic model (i.~e.,
well-defined at short distances), using the standard $T$-matrix
approach on a lattice, adopted by many authors.\cite{%
Peres2006,Pogorelov,Loktev,Lichtenstein,Peres2007,Bena}
We consider the two kinds of impurities, mentioned above, modeled as
a diagonal on-site potential for a site-like impurity, and a combination
of a diagonal and an off-diagonal potentials for a bond-like impurity
(the same model was adopted in Ref.~\onlinecite{Lichtenstein}).
The scattering lengths are calculated as functions of the impurity
strengths, and~$\cut$ is obtained to be $0.5\,a$.
In agreement with the results of Ref.~\onlinecite{Lichtenstein},
the divergence of the scattering length occurs at infinite impurity
strength for a site-like impurity (corresponding to a zero-energy
state bound to a vacancy), and at finite values of the diagonal
and off-diagonal strengths for a bond-like impurity.
The resulting cross-sections as functions of the electron energy
for different impurity strengths are shown in
Figs.~\ref{fig:sigmangstrom},\ref{fig:sigma2angstrom}.

\section{Free electrons in graphene}

Graphene unit cell contains two atoms, labelled $A$~and~$B$
(Fig.~\ref{fig:lattice}). Each of them has one $\pi$-orbital, so there
are two electronic states for each point of the first Brillouin zone
(the electron spin is not considered in the present work).
The electronic energy~$\ep$
(measured from the Fermi level of the undoped graphene) vanishes
at the two Dirac points $K,K'$ with wave vectors~$\pm\vec{K}$.
Thus, there are exactly four electronic states with $\ep=0$. An
arbitrary linear combination of them is represented by a 4-component
column vector~$\psi$. Here we choose the following arrangement
of the wave function components in the column:\cite{AleinerEfetov}
\begin{equation}
\psi=\left[\begin{array}{c} \psi_{AK} \\ \psi_{BK} \\
\psi_{BK'} \\ -\psi_{AK'} \end{array}\right].
\label{Aleinerrep=}
\end{equation}
Other definitions of the column vector are possible, but they are
inconvenient for the present problem; for the discussion
see Appendix~\ref{app:reps} and Ref.~\onlinecite{thickRaman}.
Being interested in low-energy states, we
focus on states in the vicinities of the Dirac points. The wave vectors
of these states can be written as $\vec{k}=\pm\vec{K}+\vec{p}$,
where $pa\ll{1}$ (here $a\approx{1}.42\:\mbox{\AA}$ is the C--C
bond length). Equivalently, states near the Dirac points are obtained
by including a smooth position dependence $\psi(\vec{r})$,
$\vec{r}\equiv(x,y)$.

\begin{table}
\begin{tabular}[t]{|c|c|c|c|c|c|c|} \hline
$C_{6v}$ & $E$ & $C_2$ & $2C_3$ & $2C_6$ & $\sigma_{a,b,c}$ &
$\sigma_{a,b,c}'$
\\ \hline\hline $A_1$ & 1 & 1 & 1 & 1 & 1 & 1 \\ \hline $A_2$ & 1
& 1 & 1 & 1 & $-1$ & $-1$ \\ \hline $B_2$ & 1 & $-1$ & 1 & $-1$ &
$1$ & $-1$ \\ \hline $B_1$ & 1 & $-1$ & 1 & $-1$ & $-1$ & $1$ \\
\hline $E_1$ & 2 & $-2$ & $-1$ & $1$ & 0 & 0 \\ \hline $E_2$ & 2 &
2 & $-1$ & $-1$ & 0 & 0 \\ \hline\end{tabular}\hspace{1cm}
\caption{Irreducible representations of the group $C_{6v}$ and
their characters.\label{tab:C6vC3v}}
\end{table}

\begin{table}
\begin{tabular}{|c|c|c|c|c|c|c|} \hline
irrep & 
$A_1$ & $B_1$ & $A_2$ & $B_2$ & $E_1$ & $E_2$ \\ \hline
\multicolumn{7}{|c|}{valley-diagonal matrices}\\
\hline
matrix & $\unitmatrix$ & $\Lambda_z$ & $\Sigma_z$ &
$\Sigma_z\Lambda_z$ & $\Sigma_x,\,\Sigma_y$ &
$-\Sigma_y\Lambda_z,\Sigma_x\Lambda_z$ \\ \hline
\multicolumn{7}{|c|}{valley-off-diagonal matrices} \\
\multicolumn{7}{|c|}{$\qquad\quad\;
\overbrace{\qquad\qquad\quad\;}\;
\overbrace{\qquad\qquad}\;
\overbrace{\qquad\qquad\qquad\qquad\qquad\qquad\;}$} \\
\hline
matrix & $\Sigma_z\Lambda_x$ &
$\Sigma_z\Lambda_y$ & $\Lambda_x$ & $\Lambda_y$ &
$\Sigma_y\Lambda_x,-\Sigma_x\Lambda_x$ &
$\Sigma_x\Lambda_y,\Sigma_y\Lambda_y$ \\ \hline
\end{tabular}
\caption{Classification of $4\times{4}$ hermitian matrices
by irreducible representations of the $C_{6v}$~group.
Matrices joined by braces transform through each
other under translations.\label{tab:matrices}}
\end{table}

The basis in the space of $4\times{4}$ hermitian matrices is formed
by 16 generators of the $SU(4)$ group. They can
be represented as products of two mutually commuting algebras of
Pauli matrices denoted by $\Sigma_x,\Sigma_y,\Sigma_z$ and
$\Lambda_x,\Lambda_y,\Lambda_z$,\cite{McCann,AleinerEfetov} which
fixes their algebraic relations. We denote the unit $4\times{4}$ matrix
by~$\unitmatrix$, and sometimes $\Sigma_0$ or $\Lambda_0$
to make the formulas compact. In representation~(\ref{Aleinerrep=})
$\Sigma_i$~are the Pauli matrices acting within upper and lower 2-blocks
(the sublattice subspace), while $\Lambda_i$ are the Pauli matrices acting
in the ``external'' subspace of the 2-blocks (the valley subspace). For
4-column vectors which can be represented as a direct product
\begin{equation}
\left[\begin{array}{c}
x_1y_1 \\ x_2y_1 \\ x_1y_2 \\ x_2y_2
\end{array}\right]\equiv
\left[\begin{array}{c} x_1 \\ x_2 \end{array}\right]
\otimes\left[\begin{array}{c} y_1 \\ y_2 \end{array}\right]
\equiv
\left[\begin{array}{c} x_1 \\ x_2 \end{array}\right]
\otimes(y_1\phi_K+y_2\phi_{K'}),
\end{equation}
the $\Sigma$ ~matrices act on the $x$~variables, while the
$\Lambda$~matrices act on the $y$~variables. The basis in the
valley subspace is denoted by $\phi_K,\phi_{K'}$ for future
convenience.

The matrices $\Sigma_i,\Lambda_j$, $i,j=x,y,z$, and their products have
definite transformation properties under the crystal group.
The irreducible representations of $C_{6v}$~group are listed in
Table~\ref{tab:C6vC3v}
($C_n$~denotes the rotation by $2\pi/n$, $\sigma_{a,b,c}$ are the three
reflections which swap the $K$~and~$K'$ points, $\sigma_{a,b,c}'$ are
the three reflections which swap $A$~and~$B$ atoms, the fixed point of
these operations being the center of the hexagon, see
Fig.~\ref{fig:lattice}).
The correspondence between them and the matrices is given in
Table~\ref{tab:matrices}.\cite{shortRaman,Falko2007}

The unitary matrices corresponding to the symmetry operations of the
crystal can also be written in terms of $\Sigma$~and~$\Lambda$ matrices,
independently of the representation used.\cite{thickRaman} Specifically,
$U_{C_3}=e^{(2\pi{i}/3)\Sigma_z}$ is the matrix of the $C_3$~rotation,
$U_{C_2}=\Sigma_z\Lambda_x$ -- of the $C_2$~rotation,
$U_{\sigma_a'}=\Sigma_x\Lambda_z$ -- of the $\sigma_a'$~reflection,
$U_{\sigma_a}=\Sigma_y\Lambda_y$ -- of the $\sigma_a$~reflection.
The two elementary translations by the vectors $\vec{a}_1$, $\vec{a}_2$
act on the wave function as $t_{\vec{a}_{1,2}}:\psi(\vec{r})\mapsto
{e}^{\mp({2}\pi{i}/3)\Lambda_z}\psi(\vec{r}-\vec{a}_{1,2})$.
The time reversal operation is defined as $\psi\mapsto\timerev\psi^*$,
where the unitary time reversal matrix assumes the convenient form
$\timerev=\Sigma_y\Lambda_y$ in representation~(\ref{Aleinerrep=}).

Taking the leading-order term in the expansion of the band hamiltonian
in the powers $pa\ll{1}$, we describe the electrons by the Dirac hamiltonian
\begin{equation}\label{Hdirac=}
H_0=-iv\vec\Sigma\cdot\vec\nabla,
\end{equation}
where $\vec\Sigma=(\Sigma_x,\Sigma_y)$ is a two-dimensional vector,
and $v\approx{1}0^8\:\mbox{cm/s}$ is the electron velocity. The eigenstates
of the Dirac hamiltonian with a definite value of momentum are plane waves:
\begin{subequations}\begin{eqnarray}
&&\psi^{(0)}_{\vec{p}\chirality\valley}(\vec{r})=
e^{i\vec{p}\vec{r}}\psi^{(0)}_{\varphi_\vec{p}\chirality{\valley}},\\
&&\psi^{(0)}_{\varphi\chirality{\valley}}=
\frac{1}{\sqrt{2}}
\left[\begin{array}{c} \chirality{e}^{-i\varphi/2} \\
{e}^{i\varphi/2} \end{array}\right]\otimes\phi_{\valley}=
\diag_\varphi^\dagger\,\frac{1}{2}
\left[\begin{array}{c} 1+\chirality \\ 1-\chirality \end{array}\right]\otimes\phi_{\valley},
\nonumber\\ \label{psi0=}
\end{eqnarray}\label{planewave=}\end{subequations}
with the energy $\ep_{\vec{p}\chirality\valley}=\chirality{v}p$.
The index $\chirality=\pm{1}$
distinguishes between the conduction and the valence band.
The unitary matrices
$\diag_{\varphi_\vec{p}}=e^{i\Sigma_y\pi/4}e^{i\Sigma_z\varphi_{\vec{p}}/2}$,
where $\varphi_{\vec{p}}=\arctan(p_y/p_x)$ is the polar angle of the
vector~$\vec{p}$, diagonalize the Dirac hamiltonian in the momentum
representation:
$v\vec{p}\cdot\vec\Sigma=%
\diag_{\varphi_\vec{p}}^\dagger{vp}\Sigma_z\diag_{\varphi_\vec{p}}$
(note that $\diag_{\varphi=2\pi}=-\diag_{\varphi=0}$).
The index $\valley=K,K'$ labels the valleys. As hamiltonian~(\ref{Hdirac=})
does not contain $\Lambda$~matrices, the valley subspace is degenerate,
so any other basis can be chosen.

Besides plane waves, we will need the wave functions of states with a
definite half-integer value of the ``total angular momentum''
$j_z=-i(\partial/\partial\varphi)+(1/2)\Sigma_z$ which also commutes with
the Dirac hamiltonian~(\ref{Hdirac=}):
\begin{equation}
\psi^{(0)}_{pj_z\chirality\valley}(\vec{r})=
\frac{1}{\sqrt{2}}
\left[\begin{array}{c}
\chirality{J}_{j_z-1/2}(pr)e^{i(j_z-1/2)(\varphi+\pi/2)} \\
 J_{j_z+1/2}(pr)e^{i(j_z+1/2)(\varphi+\pi/2)}
\end{array}\right]\otimes\phi_{\valley},
\end{equation}
where $J_m$~are Bessel functions of the first kind. If one relaxes
the condition of regularity of the wave function at $\vec{r}=0$,
the Bessel functions of the first kind~$J_m$ can be replaced by
the Bessel functions of the second kind~$Y_m$, or Hankel
functions $H_m^{(1,2)}=J_m\pm{i}Y_m$.

\begin{widetext}

\section{Resonant scattering on a short-range potential}
\label{sec:scattering}

\subsection{General definitions}

Here we use the standard expansion
in partial waves\cite{Newton} modified for the Dirac equation analogously to
Refs.~\onlinecite{Mele1984,Katsnelson2007,Novikov2007}.
For a potential $V(\vec{r})$ that falls off rapidly at distances
$r\gtrsim\range$,
the electron motion at distances $r\gg\range$ can be considered free.
The general scattering solution corresponding to the energy
$\ep=\chirality{vp}$ can be written as
\begin{eqnarray}
\psi_{\vec{p}\chirality{\valley}}(\vec{r})=
\sum_{m=-\infty}^\infty\frac{{e}^{-i(m+1/2)\varphi_\vec{p}}}{\sqrt{2}}
\left[\begin{array}{c}
\chirality{J}_m(pr)e^{im(\varphi+\pi/2)} \\ J_{m+1}(pr)e^{i(m+1)(\varphi+\pi/2)}
\end{array}\right]\otimes\phi_{\valley}
+\nonumber\\+
\sum_{m=-\infty}^\infty\frac{1}{\sqrt{2}}\left[\begin{array}{c}
\chirality{H}^{(1)}_m(pr)e^{im(\varphi+\pi/2)} \\
{H}^{(1)}_{m+1}(pr)e^{i(m+1)(\varphi+\pi/2)}
\end{array}\right]\otimes\frac{1}2\,\mathcal{F}_{m+1/2}^\chirality(\vec{p})\phi_{\valley}.
\label{scatteringsolution=}
\end{eqnarray}
The first sum represents just the incident plane wave~(\ref{planewave=}).
The second sum in Eq.~(\ref{scatteringsolution=}) represents the
outgoing scattered wave. Since a short-range potential can change
arbitrarily the structure of the state in the valley subspace, an
arbitrary $2\times{2}$ matrix $\mathcal{F}_{m+1/2}^\chirality(\vec{p})$
is introduced
(the factor 1/2 in front of $\mathcal{F}$ is introduced for convenience).%
\footnote{For the states with negative energies, $\chirality=-1$, the direction
of the electron group velocity is opposite to that of momentum. Thus, for
$\chirality=-1$ the scattered part of the wave function~(\ref{scatteringsolution=})
represents an {\em incoming} flux of electrons. However, we prefer to work
with the $T$-matrix for the chronologically ordered Green's function
$G(\vec{p},\ep)$, defined in Eq.~(\ref{Greenf=}), rather than for the retarded
Green's function $G^R(\vec{p},\ep)$, the analytical continuation of $G(\vec{p},\ep)$
from the positive semiaxis of~$\ep$ through the upper complex half-plane of~$\ep$.
For this purpose we work with the solutions~(\ref{scatteringsolution=})
which correspond to outgoing waves for electrons if $\chirality=1$ and for holes
if $\chirality=-1$.}
The $2\times{2}$ matrix~$\mathcal{F}$ can be represented as a linear
combination of $\Lambda$~matrices.

Using the asymptotic behaviour of $H_m^{(1)}(pr)$ at $pr\gg|m^2-1/4|$
for integer~$m$:
\begin{equation}
H_m^{(1)}(pr)\mathop{\sim}\limits_{pr\to\infty}
\sqrt{\frac{2/\pi}{pr}}\,e^{ipr-im\pi/2-i\pi/4},
\end{equation}
and the standard definitions of the scattering amplitude
$f_\chirality(\varphi,\varphi_\vec{p};p)$ and the scattering
matrix $\mathcal{S}_\chirality(\varphi,\varphi';p)$ in two
dimensions,\cite{LL3} where the existence of the degenerate valley
subspace is taken into account:
\begin{subequations}\begin{eqnarray}
&&\psi_{\vec{p}\chirality{\valley}}(\vec{r})\mathop{\sim}\limits_{r\to\infty}
e^{i\vec{p}\vec{r}}\psi^{(0)}_{\varphi_\vec{p}\chirality{\valley}}
+\frac{e^{ipr+i\pi/4}}{\sqrt{r}}\,
{f}_\chirality(\varphi_\vec{r},\varphi_\vec{p};p)\,
\psi^{(0)}_{\varphi_\vec{r}\chirality{\valley}},\\
&&\psi_{\vec{p}\chirality{\valley}}(\vec{r})\mathop{\sim}\limits_{r\to\infty}
\frac{e^{-ipr+i\pi/4}}
{\sqrt{2\pi{p}r}}\,2\pi\delta(\varphi_\vec{r}-\varphi_\vec{p}-\pi)\,
\psi^{(0)}_{\varphi_\vec{p}\chirality{\valley}}+
\frac{e^{ipr-i\pi/4}}{\sqrt{2\pi{p}r}}\,
\mathcal{S}_\chirality(\varphi_\vec{r},\varphi_\vec{p};p)\,
\psi^{(0)}_{\varphi_\vec{r}\chirality{\valley}},
\end{eqnarray}\end{subequations}
we relate them to the matrix $\mathcal{F}_{m+1/2}^\chirality(\vec{p})$;
calculating the probability current
$\vec{j}(\vec{r})=\psi^\dagger(\vec{r})\,v\vec\Sigma\,\psi(\vec{r})$,
we obtain the differential cross-section
$d\sigma_{\chirality{\valley\valley'}}(\varphi,\varphi';p)$ for an incident
particle with momentum $\vec{p}=(p\cos\varphi',p\sin\varphi')$ in the
valley~$\valley'$ to be scattered into the valley~$\valley$ in the direction
$\vec{n}=(\cos\varphi,\sin\varphi)$:
\begin{subequations}\begin{eqnarray}
&&f_\chirality(\varphi,\varphi';p)=\frac{\mathcal{F}^\chirality(\varphi,\varphi';p)}{i\sqrt{2\pi{p}}},\\
&&\mathcal{S}_\chirality(\varphi,\varphi';p)=\unitmatrix_{2\times{2}}2\pi\delta(\varphi-\varphi')
+\mathcal{F}^\chirality(\varphi,\varphi';p),\\
&&\mathcal{F}^\chirality(\varphi,\varphi_\vec{p};p)\equiv
\sum_{m=-\infty}^\infty\mathcal{F}_{m+1/2}^\chirality(\vec{p})\,e^{i(m+1/2)\varphi},\\
&&\frac{d\sigma_{\chirality{\valley\valley'}}(\varphi,\varphi';p)}{d\varphi}=
\left|\phi_{\valley}^\dagger\,
{f}_\chirality(\varphi,\varphi';p)\,\phi_{\valley'}\right|^2.
\end{eqnarray}\end{subequations}
Note that
$\psi^{(0)}_{\varphi+2\pi,\chirality{\valley}}=-\psi^{(0)}_{\varphi\chirality{\valley}}$;
the above definitions imply $0\leq\varphi_\vec{r}-\varphi_\vec{p}<2\pi$.
For $-2\pi\leq\varphi_\vec{r}-\varphi_\vec{p}<0$ we have to set
$\mathcal{S}_\chirality(\varphi,\varphi';p)=
-\mathcal{S}_\chirality(\varphi+2\pi,\varphi';p)$.
The scattering matrix satisfies the unitarity and reciprocity
conditions (the latter assuming the symmetry of the scattering
potential with respect to the time reversal):
\begin{subequations}\begin{eqnarray}
&&\int\limits_0^{2\pi}\frac{d\varphi}{2\pi}\,
\mathcal{S}^\dagger_\chirality(\varphi,\varphi_1;p)\,
\mathcal{S}_\chirality(\varphi,\varphi_2;p)=
\unitmatrix_{2\times{2}}2\pi\delta(\varphi_1-\varphi_2),
\label{unitarity=}\\
&&\mathcal{S}_\chirality(\varphi,\varphi';p)=
\Lambda_y\,
\mathcal{S}_\chirality^T(\varphi'+\pi,\varphi+\pi;p)\,
\Lambda_y,\label{reciprocity=}
\end{eqnarray}\label{conditions=}\end{subequations}
where $\mathcal{S}^T$~denotes the
transpose of the matrix~$\mathcal{S}$.
The scattering amplitude can be related to the $T$-matrix
$T(\vec{p},\vec{p}';\ep)$ on the mass shell,
$|\vec{p}|=|\vec{p}'|=|\ep|/v$. Starting from the exact
expression for the scattered wave function,
\begin{equation}
\psi_{\vec{p}\chirality{\valley}}(\vec{r})
=e^{i\vec{p}\vec{r}}\psi_{\varphi_\vec{p}\chirality{\valley}}^{(0)}
+\int\frac{d^2\vec{p}'}{(2\pi)^2}\,e^{i\vec{p}'\vec{r}}
G(\vec{p}',\chirality{v}p)\,T(\vec{p}',\vec{p};\chirality{v}p)\,
\psi_{\varphi_\vec{p}\chirality{\valley}}^{(0)},\quad
G(\vec{p},\ep)\equiv\frac{\ep\unitmatrix+v\vec{p}\cdot\vec\Sigma}
{\ep^2-(vp-i0^+)^2},\label{Greenf=}
\end{equation}
and taking its $pr\gg{1}$ asymptotics, we arrive at
[$\vec{n}=(\cos\varphi,\sin\varphi)$ and $\vec{n}'=(\cos\varphi',\sin\varphi')$
are unit vectors]:
\begin{equation}
\diag_\varphi\,T(p\vec{n},p\vec{n}';\chirality{v}p)\,\diag_{\varphi'}^\dagger=
\frac{iv}{p}\,\frac{1}{2}\left[\begin{array}{cc}
1+\chirality & 0 \\ 0 & 1-\chirality
\end{array}\right]
\otimes
\chirality\mathcal{F}^\chirality(\varphi,\varphi';p).
\label{Tmatrix=}
\end{equation}

\subsection{Scattering lengths}

At $r\sim\range$ the potential mixes different terms in
Eq.~(\ref{scatteringsolution=}).  The asymptotic behaviour of the
Bessel and Hankel functions at $pr\ll\sqrt{|m|+1}$ for integer~$m$
is given by
\begin{subequations}\begin{eqnarray}
&&J_m(pr)=\frac{(\sign{m})^m}{|m|!}\left(\frac{pr}2\right)^{|m|}
,\\
&&H_{m\neq{0}}^{(1)}(pr)=\frac{(\sign{m})^m|m|!}{i\pi|m|}
\left(\frac{2}{pr}\right)^{|m|}
,\\
&&H_{0}^{(1)}(pr)=1+\frac{2i}\pi\left(\ln\frac{pr}2+\gamma\right),
\end{eqnarray}\label{Besselshort=}\end{subequations}
where $\gamma=0.5772\ldots$ is the Euler-Mascheroni constant.
At this stage we make an assumption that all terms constituting the
scattered wave in Eq.~(\ref{scatteringsolution=}) should be of the
same order at $r\sim\range$, provided that
{\em their coupling is allowed by the symmetry of the scattering
potential}. Depending on the symmetry
of the scattering potential, we have to consider two cases.
(i) The potential is isotropic and $\Sigma_z$-conserving:
$V(\vec{r})=V(r)$,
$\Sigma_zV(\vec{r})\Sigma_z=V(\vec{r})$
(this case was analyzed in Refs.~\onlinecite{Mele1984,Katsnelson2007,Novikov2007}).
Then $j_z=-i(\partial/\partial\varphi)+(1/2)\Sigma_z$
is conserved, so the terms in Eq.~(\ref{scatteringsolution=}) with
different values of~$m$ are decoupled (each term with a given~$m$
corresponds to $j_z=m+1/2$). Matching the terms gives
$\mathcal{F}_{j_z}\sim(p\range)^{2|j_z|}$.
(ii)~The potential $V(\vec{r})$ is generic, so it mixes all states
with different values of~$j_z$. Matching the terms gives
$\mathcal{F}_{j_z}\sim(p\range)^{|j_z|+1/2}$.
Thus, $\mathcal{F}_{\pm{1}/2}$ are the most important terms in both
cases; moreover, as typically $\range\sim{a}$, considering the terms
with $|j_z|>1/2$ would require going beyond the Dirac hamiltonian, since
the Dirac hamiltonian itself is the leading term in the expansion in $pa$.

At $r\ll{1}/p$ we can neglect the
energy in the Dirac equation, which becomes
\begin{equation}
[-iv\vec\Sigma\cdot\vec\nabla+V(\vec{r})]\psi(\vec{r})=0.
\label{zeroDirac=}
\end{equation}
At $r\gg\range$ this equation admits solutions of the form
\begin{equation}
r^me^{im\varphi}\left[\begin{array}{c} 1 \\ 0 \end{array}\right]\otimes\phi_{\valley},
\quad
r^{-m}e^{im\varphi}\left[\begin{array}{c} 0 \\ 1 \end{array}\right]\otimes\phi_{\valley},
\end{equation}
which determine the asymptotics of different angular harmonics of the {\em four}
linearly independent zero-energy solutions.
Since non-zero angular harmonics of the incident wave in Eq.~(\ref{scatteringsolution=})
vanish at $p\to{0}$, we are interested in the solutions of Eq.~(\ref{zeroDirac=})
whose asymptotics can be writen as
\begin{subequations}\begin{eqnarray}
&&\psi_{01v}(\vec{r})\mathop{\sim}\limits_{r\to\infty}
\left[\begin{array}{c} 1 \\ 0 \end{array}\right]\otimes\phi_{\valley}
+\frac{e^{i\varphi}}{ir}
\left[\begin{array}{c} 0 \\ 1 \end{array}\right]\otimes\Lmat_{11}\phi_{\valley}
+\frac{e^{-i\varphi}}{ir}
\left[\begin{array}{c} 1 \\ 0 \end{array}\right]\otimes\Lmat_{21}\phi_{\valley}
+\sum_{m>1}O\!\left(\frac{e^{\pm{i}m\varphi}}{r^m}\right),\\
&&\psi_{02v}(\vec{r})\mathop{\sim}\limits_{r\to\infty}
\left[\begin{array}{c} 0 \\ 1 \end{array}\right]\otimes\phi_{\valley}
+\frac{e^{i\varphi}}{ir}
\left[\begin{array}{c} 0 \\ 1 \end{array}\right]\otimes\Lmat_{12}\phi_{\valley}
+\frac{e^{-i\varphi}}{ir}
\left[\begin{array}{c} 1 \\ 0 \end{array}\right]\otimes\Lmat_{22}\phi_{\valley}
+\sum_{m>1}O\!\left(\frac{e^{\pm{i}m\varphi}}{r^m}\right),
\end{eqnarray}\label{zeroasymptotics=}\end{subequations}
where each $\Lmat_{ij}$ is a $2\times{2}$ matrix in the valley
subspace, which has to be determined from the solution of the
Schr\"odinger equation at short distances
(the factor $1/i$ is introduced for convenience).
Let us associate the indices $i,j=1,2$ of the $2\times{2}$ matrices
$\Lmat_{ij}$ with the matrix sturcture in the $\Sigma$-subspace,
thus combining the four $2\times{2}$ matrices $\Lmat_{ij}$ into
a single $4\times{4}$ matrix~$\Lmat$. Then the asymptotic
behavior of any solution of Eq.~(\ref{zeroDirac=}) at $r\to\infty$
can be written as:
\begin{equation}
\psi(\vec{r})=
\psi^{(0)}+\frac{1}{ir}\,(\vec{n}\cdot\vec\Sigma)\Lmat\psi^{(0)}
+\sum_{m>1}O\!\left(\frac{e^{\pm{i}m\varphi}}{r^m}\right),
\label{boundarycond=}
\end{equation}
where $\psi^{(0)}$~is an arbitrary 4-column. This equation could
also be viewed as the boundary condition on the angular harmonics
of the scattering solution~(\ref{scatteringsolution=}) at $r\to{0}$;
being formed at short distances $r\sim\range$, this boundary condition
should not depend on~$\ep$ for $|\ep|\ll{v}/\range$.
However, due to the logarithmic divergence of Hankel function
$H_0^{(1)}(pr)$, matching of wave functions should be performed
at some $r=\cut\sim\range$.  Note that in contrast to the
two-dimensional Schr\"odinger equation, the value of the constant~$\cut$
cannot be determined from the zero-energy solution since the logarithmic
function is not a solution of the Dirac equation at zero energy.

Comparing expressions~(\ref{zeroasymptotics=}) to
the scattering solution~(\ref{scatteringsolution=}) and using asymptotic
expressions~(\ref{Besselshort=}),
we obtain the general possible form of
$\mathcal{F}^\chirality_{\pm{1}/2}(\vec{p})$:
\begin{subequations}\begin{eqnarray}
&&\frac{i}{\pi{p}}\,\mathcal{F}_{+1/2}^\chirality(\vec{p})=
\chirality\Lmat_{11}\left[e^{-i\varphi_\vec{p}/2}
+\frac{1}2\,H_0^{(1)}(p\cut)\,\mathcal{F}_{+1/2}^\chirality(\vec{p})\right]
+\Lmat_{12}\left[e^{i\varphi_\vec{p}/2}
+\frac{1}2\,H_0^{(1)}(p\cut)\,\mathcal{F}_{-1/2}^\chirality(\vec{p})\right],\\
&&\frac{i}{\pi{p}}\,\mathcal{F}_{-1/2}^\chirality(\vec{p})=
\Lmat_{21}\left[e^{-i\varphi_\vec{p}/2}
+\frac{1}2\,H_0^{(1)}(p\cut)\,\mathcal{F}_{+1/2}^\chirality(\vec{p})\right]
+\chirality\Lmat_{22}\left[e^{i\varphi_\vec{p}/2}
+\frac{1}2\,H_0^{(1)}(p\cut)\,\mathcal{F}_{-1/2}^\chirality(\vec{p})\right].
\label{resFmatrix=}
\end{eqnarray}\end{subequations}
\end{widetext}
Solving these equations and comparing the result to Eq.~(\ref{Tmatrix=}),
we obtain the general low-energy $T$-matrix:\footnote{
Expression~(\ref{Tmatrixres=}) can be compared to analogous
expressions for the case of scalar particles with parabolic spectrum
determined by the mass~$m$ in the two- and three-dimensional
cases
[L.~D.~Landau and E.~M.~Lifshitz,
{\it Quantum mechanics, non-relativistic theory},
(Pergamon Press, New York, 1977)]:
$T_{3D}(\ep)=(2\pi{l}/m)/(1+ipl)$, $l$~being the scattering length,
$p=\sqrt{2m\ep}$, and $T_{2D}(\ep)=(\pi/m)/[i\pi/2-\ln(e^\gamma{p}\cut/2)]$.
}
\begin{equation}
T(\ep)=\left[\unitmatrix+\frac{\ep}v
\left(\ln\frac{2v}{\cut|\ep|}-\gamma+\frac{i\pi}2\right)\Lmat\right]^{-1}
2\pi{v}\Lmat,
\label{Tmatrixres=}
\end{equation}
which determines the differential cross-section:
\begin{equation}
\frac{d\sigma_{\chirality{\valley\valley'}}(\varphi,\varphi';p)}{d\varphi}=
\frac{p}{2\pi{v}^2}\left|(\psi^{(0)}_{\varphi\chirality{\valley}})^\dagger\,
T(\chirality{v}p)\,\psi^{(0)}_{\varphi'\chirality{\valley}'}\right|^2.
\label{sigmaT=}
\end{equation}
The matrix~$\Lmat$ satisfies
(i)~$\Lmat=\Lmat^\dagger$ to ensure the unitarity
of the scattering matrix~(\ref{unitarity=}), and
(ii)~$\Lmat=\timerev\Lmat^T\timerev^\dagger$
as a consequence of the reciprocity condition~(\ref{reciprocity=}).
Thus, it has four orthogonal eigenvectors~$\psi_i$: $\Lmat\psi_i=l_i\psi_i$,
$i=1,\ldots,4$, and the four eigenvalues~$l_i$ play the role of the scattering lengths.
The angular dependence of
$d\sigma_{\chirality{\valley\valley'}}(\varphi,\varphi';p)/d\varphi$
is given by the sum of an isotropic term and terms
$\propto{e}^{\pm{i}\varphi},{e}^{\pm{i}\varphi'}$, and
${e}^{\pm{i}\varphi\pm{i}\varphi'}$
(with all four combinations of the signs).
The total out-scattering cross-section (i.~e., integrated over~$\varphi$
and summed over~$v$) takes the simple form:
\begin{equation}
\sigma_{\chirality{\valley}'}^\mathrm{out}(\varphi';p)=\sum_{i=1}^4
\frac{2\pi^2p\,|\psi_i^\dagger\psi^{(0)}_{\varphi'\chirality{\valley}'}|^2}
{\left[l_i^{-1}-\chirality{p}\ln(e^\gamma{p}\cut/2)\right]^2+(\pi{p}/2)^2}.
\label{sigmares=}
\end{equation}

Typically, one assumes that for a strong potential all scattering lengths
$l_i\sim\range$. However, an explicit calculation for a point defect in the
tight-binding model, performed in Sec.~\ref{sec:tight}, shows that one of
the lengths~$l_i$ (let it be~$l_1$ for definiteness) can become arbitrarily
large. The case $l_1\to\infty$  corresponds to the existence of a localized
solution $\psi(\vec{r})\sim(\vec{n}\cdot\vec\Sigma)\psi_1/r$ at zero energy.%
\cite{Pereira2006}
In this case the cross-section diverges at $p\to{0}$.
This divergence corresponds to a similar divergence in the imaginary part
of the electron self-energy found in Refs.~\onlinecite{Peres2006,Mirlin};
a similar divergence in the cross-section was found in
Ref.~\onlinecite{Stauber2007}.
%
Note that even in the case of resonant scattering the scaling
$\mathcal{F}_{j_z}\sim(p\range)^{|j_z|}\mathcal{F}_{\pm{1/2}}$ holds:
indeed, at $l_1\to\infty$ the coefficients at the $1/r^m$ terms in the
wave function of the localized state should scale as~$\range^m$, as
there is no other length scale in the problem.


\section{Impurities with special symmetries}\label{sec:special}

Let us consider two particular kinds of impurities, shown in Fig.~\ref{fig:lattice}.

(i) A site-like impurity with the symmetry $C_{3v}$ whose fixed point
is located on one of the atoms (let us assume it to be an
$A$~atom). Thus, the matrix $\Lmat$ should be invariant under
the reflection $\sigma_a$ and the rotation $C_3'=C_3t_{\vec{a}_1}$
(we remind that the rotation~$C_3$ is around the center of the hexagon).
The conditions
$\Lmat=U_{\sigma_a}^\dagger\Lmat{U}_{\sigma_a}$,
$\Lmat=U_{C_3'}^\dagger\Lmat{U}_{C_3'}$ together with
the time-reversal symmetry restrict the matrix $\Lmat$ to
\begin{equation}
\Lmat=
\Lmat_{A_1}\unitmatrix+\Lmat_{B_2}\Lambda_z\Sigma_z
+\tilde{\Lmat}_{E}(\Lambda_x\Sigma_x-\Lambda_y\Sigma_y).
\label{Lsite=}
\end{equation}
The eigenvalues and eigenvectors of this family of matrices are
\begin{subequations}\begin{eqnarray}
&&l_{1,2}=\Lmat_{A_1}+\Lmat_{B_2}
\pm{2}\tilde{\Lmat}_{E},\quad
l_{3,4}=\Lmat_{A_1}-\Lmat_{B_2},\\
&&\psi_{1,2}=\frac{1}{\sqrt{2}}\left[\begin{array}{c} 1 \\ 0 \\ 0 \\ \pm{1} \end{array}\right],\quad
\psi_{3,4}=\frac{1}{\sqrt{2}}\left[\begin{array}{c} 0 \\ 1 \\ \pm{1} \\ 0 \end{array}\right].
\end{eqnarray}\end{subequations}
The $T$-matrix has the same transformation properties as the
matrix~$\Lmat$, so it can be written in the same
form~(\ref{Lsite=}) with the substitution $\Lmat\to{T}$.
Then, according to Eq.~(\ref{sigmaT=}) the differential intravalley
and intervalley cross-sections can be written as
\begin{subequations}\begin{eqnarray}
&&\frac{d\sigma_{KK}}{d\varphi}=
\frac{p}{2\pi{v}^2}\left|T_{A_1}\cos\frac{\varphi-\varphi'}2
+iT_{B_2}\sin\frac{\varphi-\varphi'}2\right|^2,\nonumber\\&&\\
&&\frac{d\sigma_{K'K}}{d\varphi}=
\frac{p}{2\pi{v}^2}\,|\tilde{T}_{E}|^2.
\end{eqnarray}\end{subequations}

(ii) A bond-like impurity with the symmetry $C_{2v}$ whose fixed point is
located at the center of a bond (let us assume it to be a bond connecting
the two atoms within the same unit cell). Then, the matrix $\Lmat$
should be invariant under the reflection $\sigma_a'$ and the rotation
$C_2'=C_2t_{\vec{a}_1}t_{\vec{a}_2}$, which fixes
\begin{subequations}\begin{eqnarray}
&&\Lmat=
\Lmat_{A_1}\unitmatrix+\Lmat_{E_2}\Lambda_z\Sigma_x
+\tilde{\Lmat}_{A_1}\Lambda_x\Sigma_z
+\tilde{\Lmat}_{E_2}\Lambda_y\Sigma_y,\nonumber\\
&&\label{Lbond=}\\
&&l_{1,2}=(\Lmat_{A_1}+\tilde{\Lmat}_{E_2})
\pm(\Lmat_{E_2}+\tilde{\Lmat}_{A_1}),\\
&&l_{3,4}=(\Lmat_{A_1}-\tilde{\Lmat}_{E_2})
\pm(\Lmat_{E_2}-\tilde{\Lmat}_{A_1}),\\
&&\psi_{1,2}=\frac{1}{2}\left[\begin{array}{c} \pm{1} \\ 1 \\ 1 \\ \mp{1} \end{array}\right],\quad
\psi_{3,4}=\frac{1}{2}\left[\begin{array}{c}  1 \\ \pm{1} \\ \mp{1} \\ 1 \end{array}\right].
\end{eqnarray}\end{subequations}
The differential intravalley and intervalley cross-sections are
\begin{subequations}\begin{eqnarray}
&&\frac{d\sigma_{KK}}{d\varphi}=
\frac{p}{2\pi{v}^2}\left|T_{A_1}\cos\frac{\varphi-\varphi'}2
+\chirality T_{E_2}\cos\frac{\varphi+\varphi'}2\right|^2,\nonumber\\&&\\
&&\frac{d\sigma_{K'K}}{d\varphi}=
\frac{p}{2\pi{v}^2}\left|\tilde{T}_{A_1}\sin\frac{\varphi-\varphi'}2
+\chirality\tilde{T}_{E_2}\sin\frac{\varphi+\varphi'}2\right|^2.\nonumber\\&&
\end{eqnarray}\end{subequations}

If the location of the impurity is different from what we have assumed,
its $\Lmat$ matrix can be obtained by applying the corresponding
symmetry operation. For example, in case~(i) the matrix for an impurity
located on a $B$~atom is obtained by a $C_2$~rotation:
$\Lmat\to\Lambda_x\Sigma_z\Lmat\Lambda_x\Sigma_z$.
Obviously, these two locations can occur with equal probability, so one
could average over them. This procedure is described in the next section.

\section{Averaging over the impurities}\label{sec:averaging}

As discussed in the end of the precedeing section, the presence of defects
characterized by a certain $T$-matrix $T(\vec{p},\vec{p}',\ep)$ implies
the presence of the same (on average) number of defects of the same type,
placed in different locations with different orientations, and thus having
different $T$-matrices, but equivalent with respect to the symmetry of the
crystal.
If one studies effects which do not involve coherent scattering on several
impurities, it is sufficient to average any observable
$\mathcal{O}[T(\vec{p},\vec{p}',\ep)]$, calculated for a single impurity,
according to
\begin{eqnarray}
\overline{\mathcal{O}}=\frac{1}{3|C_{6v}|}\sum_{R\in{C}_{6v}}
\left(\mathcal{O}\!\left[U_R\,T(R\vec{p},R\vec{p}',\ep)\,U_R^\dagger\right]\right.+\nonumber\\
{}+
\mathcal{O}\!\left[U_{t_{\vec{a}_1}}U_R\,T(R\vec{p},R\vec{p}',\ep)\,
U_R^\dagger U_{t_{\vec{a}_1}}^\dagger\right]+\nonumber\\
{}+
\left.\mathcal{O}\!\left[U_{t_{\vec{a}_1}}^\dagger
U_R\,T(R\vec{p},R\vec{p}',\ep)\,
U_R^\dagger U_{t_{\vec{a}_1}}\right]\right).
\label{average=}
\end{eqnarray}
Here $|C_{6v}|=12$ is the number of elements in the $C_{6v}$~group,
$R$~are the operations from the group, and $U_R$ are their $4\times{4}$
matrices in the $\psi$-representation. The averaging is performed also
over the elementary translations with the matrices
$U_{t_{\vec{a}_1}}=e^{-(2\pi{i}/3)\Lambda_z}$ and
$U_{t_{\vec{a}_2}}=U_{t_{\vec{a}_1}}^\dagger$. It is convenient
to consider the group $C_{6v}''$ -- the direct product of the point
group~$C_{6v}$ and the 3-cyclic group represented by the matrices
$\unitmatrix,{e}^{\pm(2\pi{i}/3)\Lambda_z}$. Then Eq.~(\ref{average=})
describes simply the average over the group~$C_{6v}''$.

Let us apply this procedure to the differential cross-section.
We write the averaged Eq.~(\ref{sigmaT=}) as
\begin{equation}
\overline{\frac{d\sigma}{d\varphi}}=
\frac{p}{2\pi{v}^2}\sum_{R\in{C}_{6v}''}
\frac{\Tr\{U_R^\dagger(\psi'{\psi'}^\dagger)
U_RT^\dagger U_R^\dagger(\psi{\psi}^\dagger)U_RT\}}{|C_{6v}''|},
\label{sigmaTav=}
\end{equation}
where we abbreviated
$\psi=\psi^{(0)}_{\varphi\chirality{\valley}}$,
$\psi'=\psi^{(0)}_{\varphi'\chirality{\valley}'}$, and $T=T(\chirality{v}p)$.
Thus, equivalently, we can calculate the average
$\overline{(\psi'{\psi'}^\dagger)\otimes(\psi\psi^\dagger)}$.
In the matrices $\psi\psi^\dagger$ and $\psi'{\psi'}^\dagger$ we separate
the components corresponding to different irreducible representations
of~$C_{6v}$:
\begin{eqnarray}
\psi\psi^\dagger&=&
\frac{1}4\sum_{i,j=0,x,y,z}
\left(\psi^\dagger\Lambda_i\Sigma_j\psi\right)\Lambda_i\Sigma_j=\nonumber\\
&=&\frac{1}4\,(\unitmatrix\pm\Lambda_z)(\unitmatrix+\chirality\vec{n}\cdot\vec\Sigma),
\label{psipsiexpand=}
\end{eqnarray}
where the plus (minus) sign should be taken for $\phi_{\valley}=\phi_K$
($\phi_{\valley}=\phi_{K'}$), and $\vec{n}=(\cos\varphi,\sin\varphi)$.
Let us label the matrices $\Lambda_i\Sigma_j$ belonging to an
irreducible representation~$\rep$ of the dimensionality~$d_\rep$
as $(\Lambda\Sigma)^\rep_\ell$, where the index~$\ell=1,\ldots,d_\rep$
labels the matrices within the representation.
Then in each representation we can define the $d_\rep\times{d}_\rep$
matrices $(U_R^\rep)_{\ell\ell'}$ as
\begin{equation}
U_R^\dagger(\Lambda\Sigma)^\rep_\ell{U}_R
=\sum_{\ell'=1}^{d_\rep}(U_R^\rep)_{\ell\ell'}
(\Lambda\Sigma)^\rep_{\ell'}.\label{matrixrep=}
\end{equation}
From the orthogonality relation for the representation matrices\cite{grouptheory}
\begin{equation}
\frac{1}{|C_{6v}|}\sum_{R\in{C}_{6v}}
(U^\rep_R)^*_{\ell_1\ell_2}(U^{\rep'}_R)_{\ell_3\ell_4}=
\frac{\delta_{\rep\rep'}}{d_\rep}\,
\delta_{\ell_1\ell_3}\delta_{\ell_2\ell_4},
\end{equation}
we obtain the general expression:
\begin{widetext}\begin{eqnarray}
\overline{\frac{d\sigma_{\chirality{\valley\valley'}}(\varphi,\varphi';p)}{d\varphi}}&=&
\frac{p}{32\pi{v}^2}\left[
\Tr\left\{\Lambda_0T^\dagger(\chirality{v}p)\Lambda_0T(\chirality{v}p)\right\}+
\frac{1}2\cos(\varphi-\varphi')\sum_{j=x,y}
\Tr\left\{\Lambda_0\Sigma_jT^\dagger(\chirality{v}p)\Lambda_0\Sigma_jT(\chirality{v}p)\right\}\right]+
\nonumber\\
&&{}+(2\delta_{\kappa\kappa'}-1)\left[\Lambda_0\to\Lambda_z\right].\label{sigmaAv=}
\end{eqnarray}\end{widetext}
(here we denoted the $4\times{4}$ unit matrix~$\unitmatrix$ by~$\Lambda_0$).

To conclude this section, we note that the averaging procedure
described above is equivalent to averaging the cross-section
$d\sigma(\varphi,\varphi')/d\varphi$ over $(\varphi+\varphi')/2$
keeping $\varphi-\varphi'$ fixed (due to the symmetry of the
crystal with respect to $C_3$~rotations), and subsequent averaging
over the sign of $\varphi-\varphi'$ (due to the symmetry with respect
to reflections).

\section{Impurities in the tight-binding model}\label{sec:tight}

\subsection{Green's function}

Let us  consider the tight-binding model with nearest-neighbor
coupling as an exactly solvable example of a microscopic model
(i.~e., well-defined at short distances). The only parameter of the
clean hamiltonian is the nearest-neighbor matrix element which
we write as $-2v/(3a)$, thus expressing it in terms of the electron
velocity at the Dirac point. It is convenient to work with a
2-component wave function $\{\Psi_A(\vec{r}_n),\Psi_B(\vec{r}_n)\}$
(corresponding to the two atoms in the unit cell), where the position
of the unit cell $\vec{r}_n=n_1\vec{a}_1+n_2\vec{a}_2$ is labelled
by two integers  $n_1,n_2$.
The tight-binding hamiltonian $\mathcal{H}_0(\vec{r}_n-\vec{r}_{n'})$
is a $2\times{2}$ matrix in the sublattice space.

The scattering problem in the tight-binding model with a few-site
potential~$\mathcal{U}$
is conveniently solved using Lippmann-Schwinger equation:
\begin{equation}
\Psi=\Psi^{(0)}+\mathcal{G}(\ep)\,\mathcal{U}\Psi,
\end{equation}
where $\Psi^{(0)}$ is the incident wave, $\Psi$~is the sought wave function,
and $\mathcal{G}(\ep)=(\ep-\mathcal{H}_0)^{-1}$ is the Green's function,
explicitly given by
\begin{eqnarray}
&&\mathcal{G}(\vec{r}_n-\vec{r}_{n'},\ep)=
\int\frac{d^2\vec{k}}{A_{BZ}}\,
\frac{e^{i\vec{k}(\vec{r}_n-\vec{r}_{n'})}}{\ep^2-|t_\vec{k}|^2}
\left(\begin{array}{cc} \ep & -t_\vec{k} \\ -t^*_\vec{k} & \ep \end{array}\right),\nonumber\\
&&\\
&&t_\vec{k}=\frac{2v}{3a}
\left(1+e^{-i\vec{k}\vec{a}_1}+e^{-i\vec{k}\vec{a}_2}\right),\nonumber\\
&&A_{BZ}\equiv\frac{(2\pi)^2}{A_\mathrm{uc}}=\frac{(2\pi)^2}{\sqrt{27}a^2/2}.\nonumber
\end{eqnarray}
The large-distance behavior of $\mathcal{G}(\vec{r},\ep)$ is determined by the
singularities of the denominator, i.~e., vicinities of the Dirac points $\pm\vec{K}$,
$\vec{k}=\pm\vec{K}+\vec{p}$, where we can approximate
\begin{equation}
1+e^{-i(\pm\vec{K}+\vec{p})\vec{a}_1}+e^{-i(\pm\vec{K}+\vec{p})\vec{a}_2}
\approx\frac{3a}2\,(\mp{p}_x+ip_y).
\end{equation}
Focusing at $|\ep|\ll{v}/a$, we obtain for $r\gg{a}$
($\cc$ stands for the complex conjugate):
\begin{subequations}\begin{eqnarray}
\mathcal{G}_{AA}(\vec{r},\ep)&=&\frac{A_\mathrm{uc}|\ep|}{4i{v}^2}
\left(e^{i\vec{K}\vec{r}}+e^{-i\vec{K}\vec{r}}\right)
\chirality{H}_0^{(1)}(|\ep|r/v),\nonumber\\&&\\
\mathcal{G}_{BA}(\vec{r},\ep)&=&\frac{A_\mathrm{uc}|\ep|}{4i{v}^2}
\left(e^{i\vec{K}\vec{r}+i\varphi+i\pi/2}+\cc 
\right){H}_1^{(1)}(|\ep|r/v),\nonumber\\&&\\
\mathcal{G}_{BA}(\vec{r},0)&=&
\frac{A_\mathrm{uc}}{v}\,
\frac{e^{i\vec{K}\vec{r}+i\varphi}-e^{-i\vec{K}\vec{r}-i\varphi}}{2\pi{ir}}.
\label{boundstate=}
\end{eqnarray}\end{subequations}
We also need the Green's function at coinciding points:
\begin{subequations}\begin{eqnarray}
\mathcal{G}_{AA}(\vec{0},\ep)&=&-\frac{A_\mathrm{uc}\ep}{\pi{v}^2}
\left(\ln\frac{2v}{|\ep|\cut}-\gamma+\frac{i\pi}2\right)+O(\ep^3),\nonumber\\
&&\label{GAA0=}\\
\mathcal{G}_{BA}(\vec{0},\ep)&=&\frac{a}{2v}+O(\ep^2).\label{GBA0=}
\end{eqnarray}\end{subequations}
The value of~$\cut$ in Eq.~(\ref{GAA0=}) is determined by the integration
over the whole first Brillouin zone; numerical integration gives $e^\gamma\cut=a$
within the numerical precision.
The leading term in Eq.~(\ref{GBA0=}) can be easily obtained in the coordinate
representation using the fact that $\mathcal{H}_0^{-1}$, just like~$\mathcal{H}_0$,
is invariant under $C_3$~rotations around each carbon atom.

\subsection{One-site impurity}

Let us add the on-site potential~$U_0$ different from zero only on the
$A$~atom of the $n_1=n_2=0$ unit cell. The limit $U_0\to\infty$ is equivalent
to imposing the boundary condition $\Psi_A(\vec{0})=0$ and thus
describes a vacancy.

First of all, we note that the two plane wave states with
$\Psi_A(\vec{r}_n)=0$,
$\Psi_B(\vec{r}_n)=e^{\pm{i}\vec{K}\vec{r}_n}$
remain the zero-energy eigenstates of the hamiltonian even in the presence
of the potential.
In the representation~(\ref{Aleinerrep=}) these two states are represented
by the 4-columns
\[
\left[\begin{array}{c} 0 \\ 1 \\ 0 \\ 0 \end{array}\right]=
\left[\begin{array}{c} 0 \\ 1 \end{array}\right]\otimes\phi_K,\quad
\left[\begin{array}{c} 0 \\ 0 \\ 1 \\ 0 \end{array}\right]=
\left[\begin{array}{c} 1 \\ 0 \end{array}\right]\otimes\phi_{K'}.
\]
Comparing them to the asymptotic forms~(\ref{zeroasymptotics=}), we
see that $\Lmat_{11}\phi_{K'}=\Lmat_{21}\phi_{K'}=0$,
$\Lmat_{12}\phi_{K}=\Lmat_{22}\phi_{K}=0$.

The other two zero-energy solutions correspond to the incident wave
on the $A$-sublattice,
$\Psi_A^{(0)}(\vec{r}_n)=e^{\pm{i}\vec{K}\vec{r}_n}$,
$\Psi_B^{(0)}(\vec{r}_n)=0$.
In the representation~(\ref{Aleinerrep=}) these two states are represented
by the 4-columns
\[
\left[\begin{array}{c} 1 \\ 0 \\ 0 \\ 0 \end{array}\right]=
\left[\begin{array}{c} 1 \\ 0 \end{array}\right]\otimes\phi_K,\quad
\left[\begin{array}{c} 0 \\ 0 \\ 0 \\ -1 \end{array}\right]=
\left[\begin{array}{c} 0 \\ -1 \end{array}\right]\otimes\phi_{K'}.
\]
As the potential $U_0$ is localized on one atom, and
$\mathcal{G}(\ep=0)$ is off-diagonal in the sublattices,
the Lippmann-Schwinger equation is straightforwardly solved
to give the wave function:
\begin{equation}
\Psi_A(\vec{r}_n)=e^{\pm{i}\vec{K}\vec{r}_n},\quad
\Psi_B(\vec{r}_n)=U_0\mathcal{G}_{BA}(\vec{r}_n,\ep=0).
\end{equation}
Using Eq.~(\ref{boundstate=}), we obtain
\begin{equation}
\Lmat=\frac{A_\mathrm{uc}U_0}{\pi{v}}\,
\frac{\unitmatrix+\Lambda_z\Sigma_z+\Lambda_y\Sigma_y-\Lambda_x\Sigma_x}{4},
\end{equation}
in agreement with Eq.~(\ref{Lsite=}).
The eigenvalues of this matrix are easily found to be
$l_1=A_\mathrm{uc}U_0/(\pi{v})$,
$l_2=l_3=l_4=0$. At $U_0\to\infty$ the scattering length $l_1$ diverges.
In this case the amplitude of the incident wave can be set to zero, and
$\Psi_B(\vec{r}_n)\propto\mathcal{G}_{BA}(\vec{r},\ep=0)$
is the wave function of the state, localized on the vacancy.

At $\ep\neq{0}$ the Lippmann-Schwinger equation is solved
self-consistently for $\Psi_A(\vec{0})$ to give the wave functions:
\begin{eqnarray}
\Psi_{\vec{k}\chirality}(\vec{r}_n)&=&\left(\begin{array}{c} e^{i\Phi_{\vec{k}}/2}\\
-\chirality{e}^{-i\Phi_{\vec{k}}/2} \end{array}\right)
\frac{e^{{i}\vec{k}\vec{r}_n}}{\sqrt{2}}+\nonumber\\
&&{}+\frac{e^{i\Phi_{\vec{k}}/2}/\sqrt{2}}{U_0^{-1}-\mathcal{G}_{AA}(\vec{0},\ep)}
\left(\begin{array}{c} \mathcal{G}_{AA}(\vec{r}_n,\ep)\\
\mathcal{G}_{BA}(\vec{r}_n,\ep) \end{array}\right),
\end{eqnarray}
where $e^{i\Phi_\vec{k}}=t_\vec{k}/|t_\vec{k}|$.
This corresponds to the $T$-matrix
\begin{equation}
T(\ep)=\frac{A_\mathrm{uc}}
{U_0^{-1}-\mathcal{G}_{AA}(\vec{0},\ep)}
\frac{\unitmatrix+\Lambda_z\Sigma_z+\Lambda_y\Sigma_y-\Lambda_x\Sigma_x}{2}.
\label{Tmatrixtb=}
\end{equation}
Using Eq.~(\ref{GAA0=}), we arrive at Eq.~(\ref{Tmatrixres=}).

\begin{figure}
\includegraphics[width=8cm]{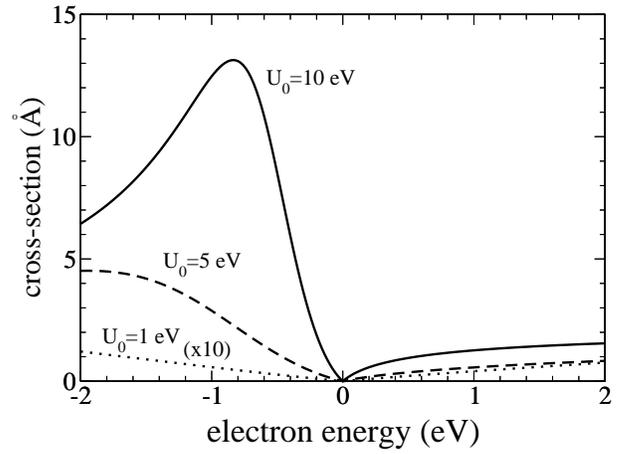}
\caption{\label{fig:sigmangstrom} The out-scattering cross-section
(coinciding with the transport cross-section) for a one-site impurity of
the strength
$U_0=1,5,10\:\mbox{eV}$ (dotted, dashed, and solid curve,
respectively) as a function of the electron energy (in eV),
as given by Eq.~(\ref{sigma1site=}).
The curve for $U_0=1\:\mbox{eV}$ is multiplied by a factor of 10.
The parameters of the model are $v=10^8\:\mbox{cm/s}=6.58\:\mbox{\AA}$,
$a=1.42\:\mbox{\AA}$.}
\end{figure}

To calculate the average cross-section, we note that in Eq.~(\ref{sigmaAv=})
only the first term survives, so the scattering is isotropic in space
and completely mixes the valleys. The total out-scattering cross-section
and the transport cross-section, averaged over the impurity positions,
coincide and are given by
\begin{equation}
\overline{\sigma^\mathrm{out}(\ep)}=\frac{\pi^2v|\ep|/2}
{\displaystyle\left(\frac{\pi{v}^2}{U_0A_\mathrm{uc}}
+\ep\ln\frac{2v}{e^{\gamma}\cut|\ep|}\right)^2
+\left(\frac{\pi\ep}{2}\right)^2}.
\label{sigma1site=}
\end{equation}
This cross-section is plotted in Fig.~\ref{fig:sigmangstrom} as
a function of~$\ep$ for several values of $U_0=1,5,10\:\mbox{eV}$.

\subsection{Two-site impurity}

Let us add a potential which mixes the two sites in the $n_1=n_2=0$ unit cell:
\begin{equation}
\mathcal{U}=\left(\begin{array}{cc} U_0 & U_1 \\ U_1 & U_0 \end{array}\right).
\label{doublesite=}
\end{equation}
We have chosen $U_1$ to be real in order to preserve the $A\leftrightarrow{B}$
symmetry.
The self-consistent solution of the Lippmann-Schwinger equation gives
\begin{equation}
\mathcal{U}\Psi(\vec{0})=
\left[\mathcal{U}^{-1}-\mathcal{G}(\vec{0},\ep)\right]^{-1}\Psi^{(0)}(\vec{0})
\equiv\mathcal{T}(\ep)\,\Psi^{(0)}(\vec{0}).
\end{equation}
Comparing the resulting wave function with Eq.~(\ref{boundarycond=}),
we obtain
\begin{eqnarray}\nonumber
\Lmat=\frac{A_\mathrm{uc}}{2\pi{v}}
\left[\mathcal{T}(0)\,\frac{\unitmatrix+\Lambda_z}{2}
+\Sigma_y\mathcal{T}(0)\Sigma_y\,\frac{\unitmatrix-\Lambda_z}{2}\right.-\\
-\left.i\mathcal{T}(0)\Sigma_y\,\frac{\Lambda_x+i\Lambda_y}{2}
+i\Sigma_y\mathcal{T}(0)\,\frac{\Lambda_x-i\Lambda_y}{2}
\right].
\end{eqnarray}
For the potential of the form~(\ref{doublesite=}) the scattering lengths
are obtained as [the matrix form of~$\Lmat$ is in agreement with
Eq.~(\ref{Lbond=})]:
\begin{subequations}\begin{eqnarray}
&&\mathcal{T}(0)=
\frac{(U_0^2-U_1^2)U_0}{U_0^2+[U_1+(a/2v)(U_0^2-U_1^2)]^2}\,\unitmatrix+\nonumber\\
&&\qquad{}+\frac{(U_0^2-U_1^2)(U_1+(a/2v)(U_0^2-U_1^2))}
{U_0^2-[U_1+(a/2v)(U_0^2-U_1^2)]^2}\,\Sigma_x\equiv\nonumber\\
&&\qquad\equiv\mathcal{T}_0\unitmatrix+\mathcal{T}_x\Sigma_x,\\
&&\Lmat=
\frac{A_\mathrm{uc}\mathcal{T}_0}{2\pi{v}}\,(\unitmatrix+\Lambda_y\Sigma_y)
+\frac{A_\mathrm{uc}\mathcal{T}_x}{2\pi{v}}\,(\Lambda_z\Sigma_x+\Lambda_x\Sigma_z),\nonumber\\ &&\\
&&l_{1,2}=\frac{A_\mathrm{uc}}{\pi{v}}\,
\frac{U_0\pm{U}_1}{1-(a/2v)(U_1\pm{U}_0)},\quad
l_{3,4}=0.
\end{eqnarray}\end{subequations}
The scattering lengths diverge when $U_1=2v/a\pm{U}_0$,
in agreement with the results of Ref.~\onlinecite{Lichtenstein}.

\begin{figure}
\includegraphics[width=8cm]{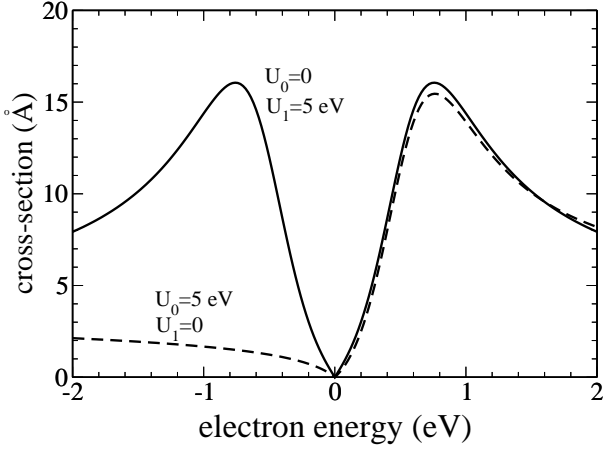}
\caption{\label{fig:sigma2angstrom} The out-scattering cross-section
for a two-site impurity with
$U_0=5\:\mbox{eV}$, $U_1=0$ (dashed curve), and
$U_0=0$, $U_1=5\:\mbox{eV}$ (solid curve)
as a function of the electron energy (in eV),
as obtained from Eq.~(\ref{sigmatwosite=}).
The parameters of the model are $v=10^8\:\mbox{cm/s}=6.58\:\mbox{\AA}$,
$a=1.42\:\mbox{\AA}$.}
\end{figure}

Calculation of the $T$-matrix from Eq.~(\ref{Tmatrixres=}) and its
substitution into Eq.~(\ref{sigmaAv=}) gives the differential
intravalley and intervalley cross-section (we set $\varphi'=0$, as it
depends only on $\varphi-\varphi'$):
\begin{subequations}\begin{eqnarray}
&&\overline{\frac{d\sigma_{\valley\valley'}}{d\varphi}}=
\frac{\pi{v}|\ep|}{8}\left[|t_1|^2+|t_2|^2\pm\frac{|t_1\pm{t}_2|^2}2\cos\varphi\right],
\label{sigmatwosite=}\\
&&t_{1,2}(\ep)=\frac{1}{v/l_{1,2}+\ep\ln[2v/(e^\gamma\cut|\ep|)]+i\pi\ep/2}.
\end{eqnarray}\end{subequations}
In Eq.~(\ref{sigmatwosite=}) the upper and lower sign is taken
for the intravalley ($\valley=\valley'$) and intervalley ($\valley\neq\valley'$)
scattering, respectively. The total out-scattering cross-section is plotted in
Fig.~\ref{fig:sigma2angstrom} for the two cases of $U_0=5\:\mbox{eV}$,
$U_1=0$, and $U_0=0$, $U_1=5\:\mbox{eV}$.

\section{Conclusions}

In this paper we have studied scattering of low-energy electrons on a single
neutral short-range impurity in graphene within the framework of the 2D Dirac
equation taking into account valley degeneracy.
We have shown that for a general short-range scatterer the most important
information needed to determine the cross-section is encoded in a $4\times{4}$
matrix~$\Lmat$, whose eigenvalues are the scattering lengths.
Divergence of one or several scattering lengths occurs whenever the impurity
has bound electronic states exactly at zero energy, which is accompanied
by the singular behavior of the scattering cross-section as a function of the
electronic energy. Quasi-bound states manifest themselves as resonances at
finite energies, their width determined by the energy itself.

The matrix~$\Lmat$ can be obtained from the solution of a
microscopic model for the impurity in graphene; only the zero-energy
states need to be considered for this. As an example of a microscopic
model we take the tight-binding model and calculate the scattering
lengths for the diagonal one-site impurity potential and the two-site
potential having both diagonal and off-diagonal components. We obtain
that one of the scattering
lengths indeed becomes much larger than the interatomic spacing for generic
strong impurities (i.~e., when impurity strength is of the order of the
electronic bandwidth). This results in (i)~a dramatic increase of the
scattering cross-section and (ii)~its strong energy dependence.

\section{Acknowledgements}

The author thanks A.~C.~Ferrari, S.~Piscanec, and F.~Guinea
for stimulating discussions.

\appendix

\section{Different representations of the state vector}
\label{app:reps}

If one adopts a representation, different from~(\ref{Aleinerrep=})
(let us denote the corresponding 4-columns by~$\tilde\psi$),
the matrices $\Sigma_i$, $\Lambda_j$, $i,j=x,y,z$, instead of being
simple Pauli matrices become some $4\times{4}$ matrices. Their
algebraic relations and symmetry
properties, listed in Table~\ref{tab:matrices}, remain the same.
In fact, the convenient way to define these matrices for an arbitrary
representation is to specify the irreducible representation of the
$C_{6v}$ group, the point group of graphene, according to which they
transform. This is sufficient to fix their algebraic
relations.\cite{thickRaman} For example, the isospin matrices
$\Sigma_x,\Sigma_y$ are defined as the matrices, diagonal
in the $K,K'$ subspace, and transforming according to the
$E_1$~representation of~$C_{6v}$.

As we have defined mutually commuting matrices $\Sigma_i$ and
$\Lambda_j$, $i,j=x,y,z$, and have written the free electron
hamiltonian in terms of the $\Sigma$~matrices only, we must separate
the degenerate valley subspace,
defined as that invariant under the action of the $\Sigma$~matrices.
Since the basis vectors of representation~(\ref{Aleinerrep=}) already have
the necessary structure of the direct product, this representation is
preferred for dealing with scattering problems. One can pass to it
by choosing four basis vectors $\tilde\psi_1,\ldots,\tilde\psi_4$,
defined as eigenvectors of $\Sigma_z$~and~$\Lambda_z$:
\begin{equation}\begin{split}
\Sigma_z\tilde\psi_1=\Lambda_z\tilde\psi_1=\tilde\psi_1,\\
-\Sigma_z\tilde\psi_2=\Lambda_z\tilde\psi_2=\tilde\psi_2,\\
\Sigma_z\tilde\psi_3=-\Lambda_z\tilde\psi_3=\tilde\psi_3,\\
-\Sigma_z\tilde\psi_4=-\Lambda_z\tilde\psi_4=\tilde\psi_4.
\end{split}\end{equation}
Their relative phases are fixed by the requirement that the
matrices $\Sigma_i$ act as the Pauli matrices in the subspaces
$\{\tilde\psi_1,\tilde\psi_2\}$ and $\{\tilde\psi_3,\tilde\psi_4\}$,
and $\Lambda_i$ act as the Pauli matrices in the subspaces
$\{\tilde\psi_1,\tilde\psi_3\}$ and $\{\tilde\psi_2,\tilde\psi_4\}$.
Thus, vectors $\psi_1,\ldots,\psi_4$ can be identified
with the basis columns $[1\,0\,0\,0]^T$,
$[0\,1\,0\,0]^T$, $[0\,0\,1\,0]^T$, $[0\,0\,0\,1]^T$
in representation~(\ref{Aleinerrep=}), up to an overall phase.
The two representations are related by a unitary matrix~$U$:
$\tilde\psi=U\psi$.

The overall phase of the matrix~$U$ is fixed by requiring the proper
form of the time reversal matrix.
In the $\tilde\psi$-representation the unitary time reversal
matrix~$\tilde\timerev$ can be different from $\Sigma_y\Lambda_y$.
Indeed, the matrices in the two representations are related by
$\tilde{U}_t=UU_tU^T$ (while $\Sigma_i$ and $\Lambda_j$ are
transformed by applying $U$~and~$U^\dagger$),
which is sensitive to the overall phase of~$U$.
However, the properties $\tilde\timerev^*\tilde\timerev=\unitmatrix$,
and $\tilde\timerev\Sigma_i^*\tilde\timerev^\dagger=-\Sigma_i$,
$\tilde\timerev\Lambda_i^*\tilde\timerev^\dagger=-\Lambda_i$,
$i=x,y,z$, do not depend on the representation. Applying these
relations in the newly constructed representation, we obtain
$U_t=\Sigma_y\Lambda_y$ up to a phase; this phase is nullified by
the appropriate choice of the phase of~$U$.


\begin{thebibliography}{99}
\bibitem{Novoselov}
K. S. Novoselov, A.~K.~Geim, S.~V.~Morozov, D.~Jiang,
Y.~Zhang, S.~V.~Dubonos, I.~V.~Grigorieva, and A.~A.~Firsov,
Science {\bf 306}, 666 (2004).
\bibitem{ripples}
J. C. Meyer, A. K. Geim, M. I. Katsnelson,
K.~S.~Novoselov, T.~J.~Booth, and S.~Roth, Nature {\bf 446}, 60 (2007).
\bibitem{Gilje2007}
S. Gilje, S.~Han, M.~Wang, K.~L.~Wang, and R.~B.~Kaner,
Nano Lett. {\bf 7}, 3394 (2007).
\bibitem{Gomez2007}
C. G\'omez-Navarro, R.~T.~Weitz, A.~M.~Bittner, M.~Scolari,
A.~Mews, M.~Burghard, and K.~Kern, Nano Lett. {\bf 7}, 3499 (2007).
\bibitem{Echtermeyer}
T. C. Echtermeyer, M. C. Lemme, M. Baus, B. N. Szafranek, A.
K. Geim, and H. Kurz, IEEE Electron Device Lett. {\bf 29}, 952
(2008).
\bibitem{Ando1998}
T. Ando and T. Nakanishi, J. Phys. Soc. Jpn. {\bf 67}, 1704 (1998).
\bibitem{Cohen2000}
H. J. Choi, J. Ihm, S. G. Louie, and M. L. Cohen,
Phys. Rev. Lett. {\bf 84}, 2917 (2000).
\bibitem{Song2002}
H.-F. Song, J.-L. Zhu, and J.-J. Xiong, Phys. Rev.~B {\bf 66}, 245421
(2002).
\bibitem{Baranger2003}
S.-H. Ke, H. U. Baranger, and W. Yang,
Phys. Rev. Lett. {\bf 91}, 116803 (2003).
\bibitem{McCann2005}
E. McCann and V. I. Fal'ko, Phys. Rev.~B {\bf 71}, 085415 (2005).
\bibitem{Pereira2006}
V.~M.~Pereira, F.~Guinea, J.~M.~B.~Lopes dos Santos, N.~M.~R.~Peres, and A.~H.~Castro Neto,
Phys. Rev. Lett. {\bf 96}, 036801 (2006).
\bibitem{Pogorelov}
Yu. G. Pogorelov, arXiv:cond-mat/0603327.
\bibitem{Peres2006}
N. M. R. Peres, F.~Guinea, and A.~H.~Castro Neto,
Phys. Rev.~B {\bf 73}, 125411 (2006).
\bibitem{Loktev}
Yu. V. Skrypnyk and V. M. Loktev, Phys. Rev.~B {\bf 73},
241402(R) (2006).
\bibitem{Lichtenstein}
T.~O.~Wehling, A.~V.~Balatsky, M.~I.~Katsnelson, A.~I.~Lichtenstein,
K.~Scharnberg, and R.~Wiesendanger,
Phys. Rev.~B {\bf 75}, 125425 (2007).
\bibitem{Peres2007}
N.~M.~R.~Peres, F.~D.~Klironomos, S.-W. Tsai, J.~R.~Santos,
J.~M.~B. Lopes dos Santos, and A.~H.~Castro Neto,
Europhys. Lett. {\bf 80}, 67007 (2007).
\bibitem{Bena}
C. Bena, Phys. Rev. Lett. {\bf 100}, 076601 (2008).
\bibitem{Yazyev}
O.~V.~Yazyev and L.~Helm, Phys. Rev. B {\bf 75}, 125408  (2007).
\bibitem{CheianovFalko}
V.~V.~Cheianov and V.~I.~Fal'ko, Phys Rev. Lett. {\bf 97}, 226801
(2006).
\bibitem{ShonAndo}
N.~H.~Shon and T.~Ando, J.~Phys. Soc. Jpn. {\bf 67}, 2421 (1998).
\bibitem{SuzuuraAndo}
H. Suzuura and T. Ando, Phys. Rev. Lett. {\bf 89}, 266603 (2002).
\bibitem{Khveshchenko}
D.~V. Khveshchenko, Phys. Rev. Lett. {\bf 97}, 036802 (2006).
\bibitem{McCann}
E. McCann, K.~Kechedzhi, V.~I.~Fal’ko, H.~Suzuura, T.~Ando, and
B.~L.~Altshuler, Phys. Rev. Lett. {\bf 97}, 146805 (2006).
\bibitem{NomuraMacDonald}
K.~Nomura and A.~H.~MacDonald,
Phys. Rev. Lett. {\bf 98}, 076602 (2007).
\bibitem{KechedzhiFalko}
K.~Kechedzhi, O.~Kashuba, and V.~I.~Fal’ko,
Phys. Rev. B 77, 193403 (2008)
\bibitem{AleinerEfetov}
I.~L.~Aleiner and K.~B.~Efetov, Phys. Rev. Lett. {\bf 97}, 236801
(2006).
\bibitem{Stauber2007}
T. Stauber, N. M. R. Peres, and F. Guinea,
Phys. Rev.~B {\bf 76}, 205423  (2007).
\bibitem{Mirlin}
P. M. Ostrovsky, I. V. Gornyi, and A. D. Mirlin,
Phys. Rev.~B {\bf 74}, 235443  (2006).
\bibitem{graphane}
J. O. Sofo, A. S. Chaudhari, and G. D. Barber,
Phys. Rev.~B {\bf 75}, 153401 (2007).
\bibitem{Newton}
R.~G.~Newton, {\it Scattering Theory of Waves and Particles},
(McGraw Hill, 1966).
\bibitem{Mele1984}
D. P. DiVincenzo and E. J. Mele, Phys. Rev.~B {\bf 29}, 1685  (1984).
\bibitem{Katsnelson2007}
M.~I.~Katsnelson and K.~S~Novoselov, Solid State Commun.
{\bf 143}, 3 (2007).
\bibitem{Novikov2007}
D.~S.~Novikov, Phys. Rev.~B {\bf 76}, 245435  (2007).
\bibitem{thickRaman}
D. M. Basko, Phys. Rev.~B {\bf 78}, 125418  (2008).
\bibitem{shortRaman}
D.~M.~Basko, Phys. Rev.~B {\bf 76}, 081405(R)  (2007).
\bibitem{Falko2007}
K. Kechedzhi, E. McCann, V.~I.~Fal'ko, H.~Suzuura, T.~Ando and B.~L.~Altshuler,
Eur. Phys.~J. Special Topics {\bf 148}, 39 (2007).
\bibitem{LL3}
L.~D.~Landau and E.~M.~Lifshitz, {\it Quantum mechanics}.
\bibitem{grouptheory}
M.~Lax, {\it Symmetry Principles in Solid State and Molecular
Physics} (Dover, New York, 2001 or Wiley, New York, 1974).
\end{thebibliography}
\end{document}